\documentclass[trackchanges]{aastex63}

\pdfoutput=1

\received{May 21, 2020}
\revised{Jul 28, 2020}
\accepted{Aug 15, 2020}
\submitjournal{ApJL}

\shorttitle{Shock waves near the Sun}
\shortauthors{Yang et al.}

\begin{document}

\title{PIC simulations of microinstabilities and waves at near-Sun solar wind perpendicular shocks: Predictions for Parker Solar Probe and Solar Orbiter}

\author[0000-0002-1509-1529]{Zhongwei Yang}
\affiliation{State Key Laboratory of Space Weather, National Space Science Center, Chinese Academy of Sciences, Beijing, 100190, People's Republic of China {\rm{\color{blue}(zwyang1984@gmail.com, liuxying@swl.ac.cn)}}}
\affiliation{CAS Key Laboratory of Geospace Environment, Chinese Academy of Sciences, University of Science and Technology of China, Hefei, 230026, People's Republic of China
{\rm{\color{blue}(qmlu@ustc.edu.cn)}}}
\affiliation{Key Laboratory of Earth and Planetary Physics, Institute of Geology and Geophysics, Chinese Academy of Sciences, Beijing, 100029, People's Republic of China}

\author[0000-0002-3483-5909]{Ying D. Liu}
\affiliation{State Key Laboratory of Space Weather, National Space Science Center, Chinese Academy of Sciences, Beijing, 100190, People's Republic of China {\rm{\color{blue}(zwyang1984@gmail.com, liuxying@swl.ac.cn)}}}
\affiliation{University of Chinese Academy of Sciences, Beijing, 100049, People's Republic of China}

\author[0000-0002-4784-0301]{Shuichi Matsukiyo}
\affiliation{Faculty of Engineering Sciences, Kyushu University, 6-1 Kasuga-Koen, Kasuga, Fukuoka, 816-8580, {\rm{\color{blue}(Japan matsukiy@esst.kyushu-u.ac.jp)}}}

\author[0000-0003-3041-2682]{Quanming Lu}
\affiliation{CAS Key Laboratory of Geospace Environment, Chinese Academy of Sciences, University of Science and Technology of China, Hefei, 230026, People's Republic of China {\rm{\color{blue}(qmlu@ustc.edu.cn)}}}

\author[0000-0003-4315-3755]{Fan Guo}
\affiliation{Los Alamos National Laboratory, NM 87545, USA {\rm{\color{blue}(guofan.ustc@gmail.com)}}}

\author[0000-0003-2981-0544]{Mingzhe Liu}
\affiliation{LESIA, Observatoire de Paris, Universit¨¦ PSL, CNRS, Sorbonne Universit¨¦, Universit¨¦ de Paris, 5 place Jules Janssen, 92195 Meudon, France {\rm{\color{blue}(mingzhe.liu@obspm.fr)}}}

\author{Huasheng Xie}
\affiliation{Hebei Key Laboratory of Compact Fusion, Langfang 065001, People's Republic of China}
\affiliation{ENN Science and Technology Development Co., Ltd., Langfang 065001, People's Republic of China
{\rm{\color{blue}(xiehuasheng@enn.cn)}}}

\author[0000-0003-0767-2267]{Xinliang Gao}
\affiliation{CAS Key Laboratory of Geospace Environment, Chinese Academy of Sciences, University of Science and Technology of China, Hefei, 230026, People's Republic of China {\rm{\color{blue}(gaoxl@mail.ustc.edu.cn)}}}

\author{Jun Guo}
\affiliation{College of Mathematics and Physics, Qingdao University of Science and Technology, Qingdao, 266061, People's Republic of China {\rm{\color{blue}(guojun@qust.edu.cn)}}}



\begin{abstract}

Microinstabilities and waves excited at moderate-Mach-number perpendicular shocks in the near-Sun solar wind are investigated by full particle-in-cell (PIC) simulations. By analyzing the dispersion relation of fluctuating field components directly issued from the shock simulation, we obtain key findings concerning wave excitations at the shock front: (1) at the leading edge of the foot, two types of electrostatic (ES) waves are observed. The relative drift of the reflected ions versus the electrons triggers an electron cyclotron drift instability (ECDI) which excites the first ES wave. Because the bulk velocity of gyro-reflected ions shifts to the direction of the shock front, the resulting ES wave propagates oblique to the shock normal. Immediately, a fraction of incident electrons are accelerated by this ES wave and a ring-like velocity distribution is generated. They can couple with the hot Maxwellian core and excite the second ES wave around the upper hybrid frequency. (2) from the middle of the foot all the way to the ramp, electrons can couple with both incident and reflected ions. ES waves excited by ECDI in different directions propagate across each other. Electromagnetic (EM) waves (X mode) emitted toward upstream are observed in both regions. They are probably induced by a small fraction of relativistic electrons. Results shed new insight on the mechanism for the occurrence of ES wave excitations and possible EM wave emissions at young CME-driven shocks in the near-Sun solar wind.

\end{abstract}




\section{Introduction} \label{sec:intro}

Collisionless shocks are of fundamental interests in astrophysics and space physics. They have been proposed as primary mechanisms for energy dissipation  \citep[e.g.,][]{Richardson2008,Parks2012} and particle acceleration \citep[e.g.,][]{Zank2006,Guo2010}. Observational studies suggest that large solar eruptions are often accompanied with shocks driven by corona mass ejections (CMEs). They are rich in various plasma waves \citep{Wilson2007,Liu2018} and usually associated with type II radio bursts \citep{Bale1999,Liu2009}. Previous investigations reveal that there are two main mechanisms for high frequency electromagnetic wave emissions (e.g., O or X modes) at collisionless shocks. One is synchrotron maser instability (SMI) which usually refers to an electromagnetic wave emission mechanism at relativistic magnetized shocks  \citep{Hoshino1991,Plotnikov2019}. This mechanism is in favor of a positive slope in a electron velocity distribution function (VDF) perpendicular to the ambient magnetic field, such as due to a loss cone or ring velocity distribution, which can become a free energy source for wave emissions. The essentially the same mechanism may operate as long as weakly relativistic anisotropic electrons exist (even down to a few $keV$) in the case of auroral electrons \citep{Wu1979}. So it may indeed be a potential candidate for radio emissions at non-relativistic CME-driven shocks associated with enhanced electron intensities from $<40keV$ to about $200keV$ \citep{liu2008}. The other is the nonlinear three-wave interaction \citep{Pulupa2010,Gao2017}. First, suprathermal and nonthermal electrons are produced in the shock. Then, these energetic electrons stimulate the growth of high-frequency electrostatic (ES) waves, such as Langmuir waves. These ES waves interact with each other in the nonuniform background solar wind or rippling shock front to produce the observed EM emissions due to the nonlinear wave interactions \citep{Umeda2010,Ganse2012}. However, it is still unclear which mechanism and what wave mode plays a major role at CME-driven shocks near the Sun.

Electrostatic waves, such as (1) Langmuir waves and their higher harmonics \citep{Bale1999,Thejappa2019} and (2) electron cyclotron harmonic waves (Bernstein waves) \citep{Wilson2010,Goodrich2018}, have been widely observed at shocks by spacecraft (ISEE 1, WIND, STEREO, CLUSTER, MMS etc.) around 1 au. They are commonly believed to accelerate electrons and provide possible free energy sources of EM emissions at shocks, magnetopause, and solar flares \citep{Graham2018,Horky2018,Henri2019}. Despite this, there is a lack of in situ observations accurately of determining shock microstructures and associated plasma waves excited in the near-Sun solar wind. \citet{Bale2016} extracted some solar wind parameters in the near-Sun conditions based on data from HELIOS and models, which give us some inspiration. The study of wave properties at shocks under the near-Sun solar wind condition could help understand ES wave excitations and EM wave emissions at young CME-driven shocks observed by Parker Solar Probe (PSP) or Solar Orbiter near the perihelion at later encounters. Near the Sun, we expect a strong magnetization and relatively low values of plasma $\beta$ \citep{Bale2016} that may affect characteristics of waves excited at the shock front. This is our motivation to do the work.

\citet{Umeda2012} directly extract the field components of the wave from a 2-D shock simulation (ambient $B_0$ in-plane case) with a relatively small box along the shock front ($\sim1c/\omega_{pi}$, where $c/\omega_{pi}$ is the ion inertial length). They identify whistler waves excited by modified two-stream instability (MTSI) \citep{Matsukiyo2003} by using a dispersion relation analysis of fluctuating electromagnetic field components $\delta B$ and $\delta E$. Based on this method, we directly extract high resolution fluctuating electromagnetic field components from 2-D large scale shock simulations (ambient $B_0$ out-of-plane case, including the complete particle gyro-motion perpendicular to the background $B_0$), and check the wave modes from the leading edge of the foot to the ramp. Furthermore, observed linear waves are confirmed by a linear theory tool (BO, a new version of PDRK) \citep{Xie2019}.

This paper is organized as follows. We present a description of the simulation model and setup in Section 2. The shock front wave analysis is presented in Section 3, and we conclude with a summary in Section 4 and discuss the implications of our results for Parker Solar Probe and Solar Orbiter.

\section{Simulation model} \label{sec:model}

We carry out shock simulations using an open source electromagnetic particle-in-cell (PIC) code named EPOCH \citep{Arber2015} with a normalization which is the same as that used in our previous works \citep{Yang2016,Yang2018,Lembege2020} to simulate the waves excited at moderate-Mach-number, perpendicular shocks. In this paper, collisionless shocks were generated by the so-called injection method \citep{Matsukiyo2012,Yang2015}, in which particles are injected from the one side of the simulation boundary (at $X=0$) at super-Alfv\'{e}nic speed $V_{inj}=6V_A$ in the $+X$ direction and specularly reflected at the other side of the simulation boundary ($X=L_x$). The shock propagates in the $-X$ direction in the present downstream rest frame. The periodic boundary condition is applied in the $Y$ direction. The number of grid cell is $n_x\times n_y=25,600\times2,000$. The spatial resolution $\Delta X=\Delta y=0.0025c/\omega_{pi}$. The box size along the shock surface is $5c/\omega_{pi}$. The ion-to-electron mass ratio $m_i/m_e$ is 100 and the particle number per cell is 50. In the 2-D simulation, the ambient magnetic field $B_0$ is along $Z$ and strictly perpendicular to the $X-Y$ simulation plane (i.e., the shock normal angle $\theta_{Bn}$) is nearly $90^\circ$. This $B_0$ configuration is similar to that in previous simulations \citep{Amano2009,Matsumoto2013}. Based on fitting methods similar to \citet{Bale2016}, plasma parameters in the near-Sun solar wind (at about $10R_s$) can be estimated by using PSP data observed around its perihelion in the first three encounters at about $36R_s$. Such as the magnetic field $B\sim591nT$, the proton density $N_{p}\sim1932cm^{-3}$, the proton temperature $T_p\sim45.3eV$, and the Alfv\'{e}nic velocity $V_A\sim290km/s$. $T_e/T_i=3$ is adopted based on previous observations from HELIOS \citep{Liu2005} and PSP \citep{Maksimovic2020}. The speed of fast CME-driven shocks at $10\ Rs$ can often exceed $1500\sim2200km/s$ \citep{Zhao2019}, and a fraction of them have extremely high speeds ($\sim3300km/s$) \citep{Liu2013,Liu2019}. Their corresponding average $\overline{M}_A$ is $5\sim9$ and the extreme ones can reach $10\sim15$. Such Mach numbers are much larger than that observed at interplanetary (IP) shocks at 1 au \citep{Wilson2007,Liu2018}. In this simulation, the Alfv\'{e}nic Mach number of the shock is about $7\sim9$. Plasma beta values: $\beta_e=0.3$ and $\beta_i=0.1$ are employed for the real solar wind condition. We adopted the upstream magnetization $\omega_{pe}/\Omega_{ce}\sim7.8$, on the order of 10 for the near-Sun solar wind conditions. In addition, we examined wave properties at 2-D shocks with similar setups for different Mach numbers: $M_A=5\sim6$ and $>10$ for slower and faster shocks, respectively. A 3-D shock simulation is also carried out for comparing the ES wave property in higher dimensions. In this paper, we focus on the moderate-Mach-number 2-D shock with universal significance. The other cases will be discussed in section 4.

\section{Simulation results: Wave analysis in 2-D shock simulations} \label{sec:results}

Figure 1a shows an overview of the time-evolving shock magnetic field $\overline{B}_z$ averaged along $Y$ in the simulation. The shock becomes mature and has reached a fully evolved state after $t>2.5\Omega_{ci}^{-1}$ in this case. In order to study wave properties at the shock front with a relatively high resolution in $\kappa-\omega$ space, the electromagnetic field components are sampled in a long period: $t=3.5\sim4.5\Omega_{ci}^{-1}$, and a large box including the whole shock front. Figure 1b-d represent the snapshots of the fluctuations of $\delta E_x$, $\delta E_y$ and $\delta B_z$ at a typical time $t=3.5\Omega_{ci}^{-1}$ within the sampling period. The zoomed $\overline{B}_z$ profile consisting typical shock structures: upstream, foot, ramp, overshoot, and downstream is shown in Figure 1d for reference. The shock ramp is located at about $X=56.4$ and marked by a vertical black line. From left to right, fluctuation profiles reveal different wave properties. Two regions are selected for wave analysis (marked by red and blue horizontal lines: A and B). In region A, ES waves are dominant and propagate in nearly the same direction. In region B, ES waves have different propagation directions and pass through each other. Long-wavelength EM waves appear and gradually increase from the middle of the foot to the ramp. Let us take the fluctuation $\delta E_x(x,y,t)$ as an example to show how the data is sampled for the wave analysis. Figure 1e-f illustrate a slice of the sampled shock profile $\delta E_x$ at $Y=L_y/2$ in regions A and B, respectively. In order to investigate the dispersion relation of waves, the 3-D fluctuation data in the $X-Y-t$ space are converted to the shock ramp rest frame. A 3-D Fourier transform is applied after Hanning windowing to compensate for the nonperiodicity of the data, in both $X$ and $t$ directions (Figure 1g-h). Similar process are also carried out for other fluctuation components $\delta E_y$ and $\delta B_z$.

Figure 2a-d shows corresponding phase space plots ($X-V_{x,y}$) of particles at $t=3.5\Omega_{ci}^{-1}$. The solar wind is coming from the left hand side and the shock ramp is marked by a black vertical line as in Figure 1b-d. In region A (Figure 2a-b), some electrons are trapped and accelerated by the excited ES waves (Figure 1b-c) at the leading edge of the foot \citep{Umeda2009,Amano2009,Yang2018}. They immediately form a ring-like velocity distribution relative to the hot drifting Maxwellian core. Figure 2c-d show that a fraction of incident ions are reflected at the ramp, and the others are directly transmitted to the downstream. At the same time, the electron bulk velocity shifts in both $X$ and $Y$ directions to keep the quasi-neutrality. The ion bulk velocity component $V_{iy}$ can be larger than $V_{ix}$ in region A due to their gyromotion and the shock acceleration along the shock surface. This overall picture is in consistent with previous simulations \citep{Scholer2003,Yang2009}. Figure 2e shows $Y-$averaged number density profiles of electrons (green), incident ions (blue), and reflected ions (red). Here, we only focus on the wave analysis region and quickly separate the reflected ions before the ramp as in \citet{Otsuka2019}. From the foot to the ramp, the percentage of reflected ions increases. Figure 2f-h represent the 2-D density profiles of electrons, reflected ions and incident ions. Combined with corresponding field fluctuations (Figure 1b-d), we find that the coupling between reflected ions and electrons is strong in region A. The coupling between incident ions and electrons along the shock normal becomes noticeable in region B.

Figure 3 shows the 3-D view of the wave dispersion relation diagram of $\delta E_x$, $\delta E_y$, and $\delta B_z$ in the $\kappa_x-\kappa_y-\omega$ space. Top and bottom panels denote results from regions A and B, respectively. In the simulation, the electron plasma frequency $\omega_{pe}$ and the Debye length $\lambda_{De}$ as units are measured in the far upstream undisturbed solar wind. By using a unified tool for plasma waves and instabilities analysis: BO \citep{Xie2019}, corresponding dispersion relations calculated by the linear theory are shown in Figure 4 to assist in identifying wave modes observed in the simulation. The input plasma parameters for the linear analysis are directly issued from the simulation data, which is averaged over the whole sampling period in the shock rest frame. More details are shown in Table 1.

Firstly, we study the waves in region A (the leading edge of the foot). In region A, three main wave modes are observed: (1) the first is an ES wave marked by ``ES-1" in Figure 3a-b. This wave is rapidly excited by the relative drift between incident electrons and gyro-reflected ions. As shown in Figure 2, the main bulk velocity component of reflected ions is in $+Y$ and $-X$ directions at the beginning of the foot. So the reflected ion beam is strongly coupled with the Doppler-shifted electron cyclotron harmonic braches on the $+\kappa_y$ and $-\kappa_x$ side (Figure 4a-b). This coupling process is discussed as electron cyclotron drift instability (ECDI) by \citet{Muschietti2013}. They assumed that the reflected ion beam is straight along the shock normal and the gyro-motion is not considered (i.e., only focus on the $X$ direction). From the simulation above, we realize that the ion bulk velocity in the $Y$ direction can be large, and even plays a more important role. This is a new point. (2) the second is also an ES wave marked by ``ES-2" in Figure 3a-b. After the ``ES-1" is excited, a fraction of incident electrons are trapped and accelerated by the ``ES-1" wave. One possible acceleration mechanism of these electrons is the shock surfing acceleration where the electric field or potential plays an important role \citep{Zank1996,Hoshino2002,Amano2009,Matsumoto2017}. The accelerated electrons immediately form a ring-like velocity distribution. This VDF has a weak positive slope in the velocity space perpendicular to the magnetic field. Free energy is quickly released through the coupling between the ring and the hot Maxwellian core. The growth rate of ``ES-2" peaks at about $\kappa=\pm0.23\lambda_{De}^{-1}$ in the linear theory around the Doppler-shifted upper hybrid frequency (Figure 4). It is consistent with the simulation results (purple regions in Figure 3a-b). (3) the third wave is an electromagnetic mode (marked by ``EM-1" in Figure 3c). This high frequency EM wave is emitted facing the upstream and is visible in both $\delta E_{\perp}$ and $\delta B_z$ diagrams. In this case, the $\delta E_{z}$ only has a background noise (not shown here). Hence, the EM mode is an extraordinary electromagnetic mode (X mode). This X mode emission cannot be described in Figure 4 because kinetic relativistic effects are not included in the present version of our linear solver yet. One possible scenario is the synchrotron maser radiation excited by relativistic electrons with a ring-type velocity distribution at strongly magnetized plasmas \citep{Hoshino1991,Plotnikov2019} under a Doppler-shifted condition. In summary, this EM wave is locally emitted at the shock front and propagates toward the upstream.

Secondly, the waves in region B (i.e., in the middle of the foot) are investigated. Figure 2e shows that the values of $n_e$ and $B$ are $\geq2$ times of their upstream values $n_{e0}$ and $B_0$ in this region. The local ratios $\omega_{pe}/\omega_{pe0}$ and $\Omega_{ce}/\Omega_{ce0}$ are about 1.414 and 2, respectively. The frequency ratio $\omega_{pe}/\Omega_{ce}$ is about 5.5 and lower than its upstream value ($\omega_{pe0}/\Omega_{ce0}\approx7.8$). This means that the magnetization of the plasma becomes greater. We keep using the upstream values $\omega_{pe0}$ amd $\lambda_{De0}^{-1}$ as units of $\omega$ and $\kappa$ for region B in bottom panels of Figure 3. Corresponding dispersion relations from the linear theory are shown in Figure 4c-d. In region B, both reflected and incident ion beams couple with the electron cyclotron harmonic branches in all directions of the simulation plane. The main excited waves are as follows: (1) in Figure 3d, ``ES-3" wave modes are excited by ECDI on $\pm\kappa_x$ directions. The wave frequency in the $\pm\kappa_x$ directions are comparable to each other as predicted by the linear theory (Figure 4c). (2) in Figure 3e, ``ES-4" wave modes have more harmonic branches in the $+\kappa_y$ direction. This is because the old reflected ions are gyrating back towards the ramp and they have a large bulk velocity ($>10V_A$) in the $+Y$ direction. In contrast, the bulk velocity of incident ions in the $-Y$ direction caused by the deflection ahead of the ramp is relatively low. In summary, the middle of the foot could be a zoom of ECDI which is triggered in different directions. This is interesting and brand new relative to previous 1-D simulations \citep{Muschietti2006,Muschietti2013} which only discuss ECDI along the $x-$axis. Our work has taken a step forward on this basis, and find that the ES waves can be excited along the shock surface as well as along the shock normal. In addition, we also find some low frequency EM waves (marked by ``EM-2" in Figure 3f) associated with the ``ES-4" waves. Such low frequency EM wave has a modulating effect on the magnetic field and plasma density profiles (refer Figure 1d and Figure 2f-h). The analysis of such EM wave amplification might requires a theory considering nonuniform plasmas and gradient background magnetic fields. Hence, it will not be further doscussed in this paper. It is worthy noting that the X mode can exist well in region B due to the local relativistic electrons, and it has a wave vector towards upstream as in region A.

\section{Conclusions and discussions} \label{sec:conclusion}

This paper presents PIC simulations of a perpendicular shock in the near-Sun solar wind condition, where the magnetization is relatively high, $\beta$ is relatively low, and the average shock Mach number is below 10, corresponding to a young, fast CME-driven shock propagating in the pristine solar wind. The simulated electromagnetic fluctuations show that the shock foot can be segmented into two parts by different wave features.

1. At the leading edge of the foot, ES waves are excited by electron cyclotron drift instability in the $-X$ and $+Y$ directions. The instability is triggered by the coupling between the incident electrons and gyro-reflected ions. The wave vector of this ES wave is oblique to the shock normal, and is mainly along the shock surface. This is because the bulk velocity of reflected ions changes from the $-X$ direction to the $+Y$ direction during the gyro-reflection process.

2. In the same region, a fraction of incident electrons can be trapped and accelerated by the above exited ES waves. A secondary instability occurs around the Doppler-shifted upper hybrid frequency due to the coupling between this accelerated ring-like electrons and the hot Maxwellian core electrons. In addition, some weak relativistic electrons in the ring VDF could lead to a high frequency EM emission probably induced by SMI. The emitted EM wave is a X mode and propagates upstream. Other potential candidates of X mode emission mechanisms are also need to be considered for quasi-perpendicular shocks. For an instance, a combination of wave growth due to electron cyclotron maser instability and nonlinear wave-coupling processes is suggested for plasmas with $\omega_{pe}/\Omega_{ce}\sim10$ in the outer corona \citep{Ni2020}.

3. From the middle of the foot all the way to the ramp, the incident ions begin to couple with the electrons. In this region, the incident ion beam begin to deflect in the $-Y$ direction when approaching the ramp, and the old gyro-back reflected ions have a large bulk velocity in the $+Y$ direction. Therefore, ES waves can be excited by ECDI in both $\pm X$ and $\pm Y$ directions. These multi-directional ES waves propagate across each other and form fibrous-like pattern in electric field and particle number density profiles. This is brand new and different from early 1-D simulations and theoretical models of shock front ES wave excitations.

4. The long-wavelength low-frequency EM waves associated with ES harmonics in the $+Y$ direction is strengthen as it approaches the ramp. In addition, the high-frequency X mode is also observed in this region.

Furthermore, we carried out two additional shock simulations with lower and higher Mach numbers as mentioned in Section 2. Preliminary results indicate that the ECDI is robust and can be observed in all cases. Buneman instability appears only at extremely fast CME-driven shocks with a speed $>3000\ km/s$ (i.e., $Ma>10$) \citep{Liu2019}. Normally this is not the case and the shock speed is as slow as that used in this paper. At slower shocks (Ma$\leq6$), the reflected ion beam is modulated by the shock front self-reformation \citep{Amano2009} and results in a intermittent ES wave excitation.

A 3-D shock simulation is also carried out for comparing ES wave excitations (not shown here). Briefly, ES waves sampled in the $B_0$ out-of-plane from the 3-D simulation is similar to that observed in this paper. However, ES waves sampled in the $B_0$ in-plane propagate almost along the shock normal. This is because the main ion bulk velocity components is not included in this cross section. More completely, a 4-D FFT is required for the wave analysis in 3-D shock cases. This is still infusible under most of the current storage conditions.

As we all know, the Sun is entering the solar maximum in the next $5\sim6$ years (solar cycle 25). More solar eruptions accompanied with fast CME-driven shocks are expected. The perihelion of PSP will be closer to the Sun (from $27.8\ R_s$ to less than $10\ R_s$) \citep{Bale2016,Fox2016,Kasper2016}. Parker Solar Probe and Solar Orbiter are likely to observe high frequency EM emissions (e.g., X modes) toward upstream accompanied with local energetic electrons, and electrostatic waves induced by ECDI or BI along the shock normal or along the shock surface at fast CME-driven shocks.

\acknowledgments

We thank the referee for helpful comments and the NASA Parker Solar Probe Mission for use of data. The authors are grateful to Dr. L. Muschietti from UC Berkeley for helpful discussions on PDRK/BO benchmarks. The computations are performed by Numerical Forecast Modeling R\&D and VR System of State Key Laboratory of Space Weather, and HPC of Chinese Meridian Project. This work is supported by the NSFC (41574140, 41674168, 41774179), the Specialized Research Fund for State Key Laboratories of China, Youth Innovation Promotion Association of the CAS (2017188), the Open Research Program Key laboratory of Geospace Environment CAS (GE2017-01), the Open Research Program Key laboratory of Polar Science, MNR (KP202005), Beijing NSF (1192018), Beijing Municipal Science and Technology Commission (Z191100004319001, Z191100004319003), Beijing Outstanding Talent Training Foundation (2017000097607G049), and the Strategic Priority Research Program of CAS (XDA14040404). S. M. acknowledges partial support by Grant-in-Aid for Scientific Research (C) No.19K03953  and (B) No.17H02966 from JSPS. J. G. is supported by the Shandong Provincial National Natural Science Foundation(ZR2017MD012).

%



\bibliography{references}{}

\begin{thebibliography}{}
\expandafter\ifx\csname natexlab\endcsname\relax\def\natexlab#1{#1}\fi
\providecommand{\url}[1]{\href{#1}{#1}}
\providecommand{\dodoi}[1]{doi:~\href{http://doi.org/#1}{\nolinkurl{#1}}}
\providecommand{\doeprint}[1]{\href{http://ascl.net/#1}{\nolinkurl{http://ascl.net/#1}}}
\providecommand{\doarXiv}[1]{\href{https://arxiv.org/abs/#1}{\nolinkurl{https://arxiv.org/abs/#1}}}

\bibitem[{{Amano} \& {Hoshino}(2009)}]{Amano2009}
{Amano}, T., \& {Hoshino}, M. 2009, \apj, 690, 244,
  \dodoi{10.1088/0004-637X/690/1/244}

\bibitem[{{Arber} {et~al.}(2015){Arber}, {Bennett}, {Brady},
  {Lawrence-Douglas}, {Ramsay}, {Sircombe}, {Gillies}, {Evans}, {Schmitz},
  {Bell}, \& {Ridgers}}]{Arber2015}
{Arber}, T.~D., {Bennett}, K., {Brady}, C.~S., {et~al.} 2015, ppcf, 57, 113001,
  \dodoi{10.1088/0741-3335/57/11/113001}

\bibitem[{{Bale} {et~al.}(1999){Bale}, {Reiner}, {Bougeret}, {Kaiser},
  {Krucker}, {Larson}, \& {Lin}}]{Bale1999}
{Bale}, S.~D., {Reiner}, M.~J., {Bougeret}, J.-L., {et~al.} 1999, \grl, 26,
  1573, \dodoi{10.1029/1999GL900293}

\bibitem[{{Bale} {et~al.}(2016){Bale}, {Goetz}, {Harvey}, {Turin}, {Bonnell},
  {Dudok de Wit}, {Ergun}, {MacDowall}, {Pulupa}, {Andre}, {Bolton},
  {Bougeret}, {Bowen}, {Burgess}, {Cattell}, {Chandran}, {Chaston}, {Chen},
  {Choi}, {Connerney}, {Cranmer}, {Diaz-Aguado}, {Donakowski}, {Drake},
  {Farrell}, {Fergeau}, {Fermin}, {Fischer}, {Fox}, {Glaser}, {Goldstein},
  {Gordon}, {Hanson}, {Harris}, {Hayes}, {Hinze}, {Hollweg}, {Horbury},
  {Howard}, {Hoxie}, {Jannet}, {Karlsson}, {Kasper}, {Kellogg}, {Kien},
  {Klimchuk}, {Krasnoselskikh}, {Krucker}, {Lynch}, {Maksimovic}, {Malaspina},
  {Marker}, {Martin}, {Martinez-Oliveros}, {McCauley}, {McComas}, {McDonald},
  {Meyer-Vernet}, {Moncuquet}, {Monson}, {Mozer}, {Murphy}, {Odom},
  {Oliverson}, {Olson}, {Parker}, {Pankow}, {Phan}, {Quataert}, {Quinn},
  {Ruplin}, {Salem}, {Seitz}, {Sheppard}, {Siy}, {Stevens}, {Summers}, {Szabo},
  {Timofeeva}, {Vaivads}, {Velli}, {Yehle}, {Werthimer}, \&
  {Wygant}}]{Bale2016}
{Bale}, S.~D., {Goetz}, K., {Harvey}, P.~R., {et~al.} 2016, \ssr, 204, 49,
  \dodoi{10.1007/s11214-016-0244-5}

\bibitem[{{Fox} {et~al.}(2016){Fox}, {Velli}, {Bale}, {Decker}, {Driesman},
  {Howard}, {Kasper}, {Kinnison}, {Kusterer}, {Lario}, {Lockwood}, {McComas},
  {Raouafi}, \& {Szabo}}]{Fox2016}
{Fox}, N.~J., {Velli}, M.~C., {Bale}, S.~D., {et~al.} 2016, \ssr, 204, 7,
  \dodoi{10.1007/s11214-015-0211-6}

\bibitem[{{Ganse} {et~al.}(2012){Ganse}, {Kilian}, {Vainio}, \&
  {Spanier}}]{Ganse2012}
{Ganse}, U., {Kilian}, P., {Vainio}, R., \& {Spanier}, F. 2012, \solphys, 280,
  551, \dodoi{10.1007/s11207-012-0077-7}

\bibitem[{{Gao} {et~al.}(2017){Gao}, {Lu}, \& {Wang}}]{Gao2017}
{Gao}, X., {Lu}, Q., \& {Wang}, S. 2017, \grl, 44, 5269,
  \dodoi{10.1002/2017GL073829}

\bibitem[{{Goodrich} {et~al.}(2018){Goodrich}, {Ergun}, {Schwartz}, {Wilson},
  {Newman}, {Wilder}, {Holmes}, {Johlander}, {Burch}, {Torbert},
  {Khotyaintsev}, {Lindqvist}, {Strangeway}, {Russell}, {Gershman}, {Giles}, \&
  {Andersson}}]{Goodrich2018}
{Goodrich}, K.~A., {Ergun}, R., {Schwartz}, S.~J., {et~al.} 2018, \jgr, 123,
  9430, \dodoi{10.1029/2018JA025830}

\bibitem[{{Graham} {et~al.}(2018){Graham}, {Vaivads}, {Khotyaintsev},
  {Andr{\'e}}, {Le Contel}, {Malaspina}, {Lindqvist}, {Wilder}, {Ergun},
  {Gershman}, {Giles}, {Magnes}, {Russell}, {Burch}, \& {Torbert}}]{Graham2018}
{Graham}, D.~B., {Vaivads}, A., {Khotyaintsev}, Y.~V., {et~al.} 2018, \jgr,
  123, 2630, \dodoi{10.1002/2017JA025034}

\bibitem[{{Guo} \& {Giacalone}(2010)}]{Guo2010}
{Guo}, F., \& {Giacalone}, J. 2010, \apj, 715, 406,
  \dodoi{10.1088/0004-637X/715/1/406}

\bibitem[{{Henri} {et~al.}(2019){Henri}, {Sgattoni}, {Briand}, {Amiranoff}, \&
  {Riconda}}]{Henri2019}
{Henri}, P., {Sgattoni}, A., {Briand}, C., {Amiranoff}, F., \& {Riconda}, C.
  2019, Journal of Geophysical Research (Space Physics), 124, 1475,
  \dodoi{10.1029/2018JA025707}

\bibitem[{{Hork{\'y}} {et~al.}(2018){Hork{\'y}}, {Omura}, \&
  {Santol{\'\i}k}}]{Horky2018}
{Hork{\'y}}, M., {Omura}, Y., \& {Santol{\'\i}k}, O. 2018, Physics of Plasmas,
  25, 042905, \dodoi{10.1063/1.5025912}

\bibitem[{{Hoshino} \& {Arons}(1991)}]{Hoshino1991}
{Hoshino}, M., \& {Arons}, J. 1991, Physics of Fluids B, 3, 818,
  \dodoi{10.1063/1.859877}

\bibitem[{{Hoshino} \& {Shimada}(2002)}]{Hoshino2002}
{Hoshino}, M., \& {Shimada}, N. 2002, \apj, 572, 880, \dodoi{10.1086/340454}

\bibitem[{{Kasper} {et~al.}(2016){Kasper}, {Abiad}, {Austin}, {Balat-Pichelin},
  {Bale}, {Belcher}, {Berg}, {Bergner}, {Berthomier}, {Bookbinder}, {Brodu},
  {Caldwell}, {Case}, {Chandran}, {Cheimets}, {Cirtain}, {Cranmer}, {Curtis},
  {Daigneau}, {Dalton}, {Dasgupta}, {DeTomaso}, {Diaz-Aguado}, {Djordjevic},
  {Donaskowski}, {Effinger}, {Florinski}, {Fox}, {Freeman}, {Gallagher},
  {Gary}, {Gauron}, {Gates}, {Goldstein}, {Golub}, {Gordon}, {Gurnee}, {Guth},
  {Halekas}, {Hatch}, {Heerikuisen}, {Ho}, {Hu}, {Johnson}, {Jordan},
  {Korreck}, {Larson}, {Lazarus}, {Li}, {Livi}, {Ludlam}, {Maksimovic},
  {McFadden}, {Marchant}, {Maruca}, {McComas}, {Messina}, {Mercer}, {Park},
  {Peddie}, {Pogorelov}, {Reinhart}, {Richardson}, {Robinson}, {Rosen},
  {Skoug}, {Slagle}, {Steinberg}, {Stevens}, {Szabo}, {Taylor}, {Tiu}, {Turin},
  {Velli}, {Webb}, {Whittlesey}, {Wright}, {Wu}, \& {Zank}}]{Kasper2016}
{Kasper}, J.~C., {Abiad}, R., {Austin}, G., {et~al.} 2016, \ssr, 204, 131,
  \dodoi{10.1007/s11214-015-0206-3}

\bibitem[{Lembege {et~al.}(2020)Lembege, Yang, \& Zank}]{Lembege2020}
Lembege, B., Yang, Z., \& Zank, G.~P. 2020, The Astrophysical Journal, 890, 48,
  \dodoi{10.3847/1538-4357/ab65c5}

\bibitem[{{Liu} {et~al.}(2018){Liu}, {Liu}, {Yang}, {Wilson}, \&
  {Hu}}]{Liu2018}
{Liu}, M., {Liu}, Y.~D., {Yang}, Z., {Wilson}, III, L.~B., \& {Hu}, H. 2018,
  \apjl, 859, L4, \dodoi{10.3847/2041-8213/aac269}

\bibitem[{{Liu} {et~al.}(2009){Liu}, {Luhmann}, {Bale}, \& {Lin}}]{Liu2009}
{Liu}, Y., {Luhmann}, J.~G., {Bale}, S.~D., \& {Lin}, R.~P. 2009, \apjl, 691,
  L151, \dodoi{10.1088/0004-637X/691/2/L151}

\bibitem[{{Liu} {et~al.}(2005){Liu}, {Richardson}, \& {Belcher}}]{Liu2005}
{Liu}, Y., {Richardson}, J.~D., \& {Belcher}, J.~W. 2005, \planss, 53, 3,
  \dodoi{10.1016/j.pss.2004.09.023}

\bibitem[{{Liu} {et~al.}(2008){Liu}, {Luhmann}, {M{\"u}ller-Mellin},
  {Schroeder}, {Wang}, {Lin}, {Bale}, {Li}, {Acu{\~n}a}, \&
  {Sauvaud}}]{liu2008}
{Liu}, Y., {Luhmann}, J.~G., {M{\"u}ller-Mellin}, R., {et~al.} 2008, \apj, 689,
  563, \dodoi{10.1086/592031}

\bibitem[{{Liu} {et~al.}(2013){Liu}, {Luhmann}, {Lugaz}, {M{\"o}stl}, {Davies},
  {Bale}, \& {Lin}}]{Liu2013}
{Liu}, Y.~D., {Luhmann}, J.~G., {Lugaz}, N., {et~al.} 2013, \apj, 769, 45,
  \dodoi{10.1088/0004-637X/769/1/45}

\bibitem[{{Liu} {et~al.}(2019){Liu}, {Zhu}, \& {Zhao}}]{Liu2019}
{Liu}, Y.~D., {Zhu}, B., \& {Zhao}, X. 2019, \apj, 871, 8,
  \dodoi{10.3847/1538-4357/aaf425}

\bibitem[{{Maksimovic} {et~al.}(2020){Maksimovic}, {Bale},
  {Ber{\v{c}}i{\v{c}}}, {Bonnell}, {Case}, {Wit}, {Goetz}, {Halekas}, {Harvey},
  {Issautier}, {Kasper}, {Korreck}, {Jagarlamudi}, {Lahmiti}, {Larson},
  {Lecacheux}, {Livi}, {MacDowall}, {Malaspina}, {Martinovi{\'c}},
  {Meyer-Vernet}, {Moncuquet}, {Pulupa}, {Salem}, {Stevens},
  {{\v{S}}tver{\'a}k}, {Velli}, \& {Whittlesey}}]{Maksimovic2020}
{Maksimovic}, M., {Bale}, S.~D., {Ber{\v{c}}i{\v{c}}}, L., {et~al.} 2020,
  \apjs, 246, 62, \dodoi{10.3847/1538-4365/ab61fc}

\bibitem[{{Matsukiyo} \& {Scholer}(2003)}]{Matsukiyo2003}
{Matsukiyo}, S., \& {Scholer}, M. 2003, \jgr, 108, 1459,
  \dodoi{10.1029/2003JA010080}

\bibitem[{{Matsukiyo} \& {Scholer}(2012)}]{Matsukiyo2012}
---. 2012, \jgr, 117, A11105, \dodoi{10.1029/2012JA017986}

\bibitem[{{Matsumoto} {et~al.}(2013){Matsumoto}, {Amano}, \&
  {Hoshino}}]{Matsumoto2013}
{Matsumoto}, Y., {Amano}, T., \& {Hoshino}, M. 2013, \prl, 111, 215003,
  \dodoi{10.1103/PhysRevLett.111.215003}

\bibitem[{{Matsumoto} {et~al.}(2017){Matsumoto}, {Amano}, {Kato}, \&
  {Hoshino}}]{Matsumoto2017}
{Matsumoto}, Y., {Amano}, T., {Kato}, T.~N., \& {Hoshino}, M. 2017, \prl, 119,
  1, \dodoi{10.1103/PhysRevLett.119.105101}

\bibitem[{{Muschietti} \& {Lemb{\`e}ge}(2006)}]{Muschietti2006}
{Muschietti}, L., \& {Lemb{\`e}ge}, B. 2006, ASR, 37, 483,
  \dodoi{10.1016/j.asr.2005.03.077}

\bibitem[{{Muschietti} \& {Lemb{\`e}ge}(2013)}]{Muschietti2013}
---. 2013, \jgr, 118, 2267, \dodoi{10.1002/jgra.50224}

\bibitem[{{Ni} {et~al.}(2020){Ni}, {Chen}, {Li}, {Zhang}, {Ning}, {Kong},
  {Wang}, \& {Hosseinpour}}]{Ni2020}
{Ni}, S., {Chen}, Y., {Li}, C., {et~al.} 2020, \apjl, 891, L25,
  \dodoi{10.3847/2041-8213/ab7750}

\bibitem[{{Otsuka} {et~al.}(2019){Otsuka}, {Matsukiyo}, \& {Hada}}]{Otsuka2019}
{Otsuka}, F., {Matsukiyo}, S., \& {Hada}, T. 2019, High Energy Density Physics,
  33, 100709, \dodoi{10.1016/j.hedp.2019.100709}

\bibitem[{{Parks} {et~al.}(2012){Parks}, {Lee}, {McCarthy}, {Goldstein}, {Fu},
  {Cao}, {Canu}, {Lin}, {Wilber}, {Dandouras}, {R{\'e}me}, \&
  {Fazakerley}}]{Parks2012}
{Parks}, G.~K., {Lee}, E., {McCarthy}, M., {et~al.} 2012, \prl, 108, 061102,
  \dodoi{10.1103/PhysRevLett.108.061102}

\bibitem[{{Plotnikov} \& {Sironi}(2019)}]{Plotnikov2019}
{Plotnikov}, I., \& {Sironi}, L. 2019, \mnras, 485, 3816,
  \dodoi{10.1093/mnras/stz640}

\bibitem[{{Pulupa} {et~al.}(2010){Pulupa}, {Bale}, \& {Kasper}}]{Pulupa2010}
{Pulupa}, M.~P., {Bale}, S.~D., \& {Kasper}, J.~C. 2010, \jgr, 115, A04106,
  \dodoi{10.1029/2009JA014680}

\bibitem[{{Richardson} {et~al.}(2008){Richardson}, {Kasper}, {Wang}, {Belcher},
  \& {Lazarus}}]{Richardson2008}
{Richardson}, J.~D., {Kasper}, J.~C., {Wang}, C., {Belcher}, J.~W., \&
  {Lazarus}, A.~J. 2008, \nat, 454, 63, \dodoi{10.1038/nature07024}

\bibitem[{{Scholer} {et~al.}(2003){Scholer}, {Shinohara}, \&
  {Matsukiyo}}]{Scholer2003}
{Scholer}, M., {Shinohara}, I., \& {Matsukiyo}, S. 2003, \jgr, 108, 1014,
  \dodoi{10.1029/2002JA009515}

\bibitem[{{Thejappa} \& {MacDowall}(2019)}]{Thejappa2019}
{Thejappa}, G., \& {MacDowall}, R.~J. 2019, \apj, 883, 199,
  \dodoi{10.3847/1538-4357/ab3bcf}

\bibitem[{{Umeda}(2010)}]{Umeda2010}
{Umeda}, T. 2010, \jgr, 115, A01204, \dodoi{10.1029/2009JA014643}

\bibitem[{{Umeda} {et~al.}(2012{\natexlab{a}}){Umeda}, {Kidani}, {Matsukiyo},
  \& {Yamazaki}}]{Umeda2012}
{Umeda}, T., {Kidani}, Y., {Matsukiyo}, S., \& {Yamazaki}, R.
  2012{\natexlab{a}}, \jgr, 117, A03206, \dodoi{10.1029/2011JA017182}

\bibitem[{{Umeda} {et~al.}(2012{\natexlab{b}}){Umeda}, {Matsukiyo}, {Amano}, \&
  {Miyoshi}}]{umeda2012b}
{Umeda}, T., {Matsukiyo}, S., {Amano}, T., \& {Miyoshi}, Y. 2012{\natexlab{b}},
  Physics of Plasmas, 19, 072107, \dodoi{10.1063/1.4736848}

\bibitem[{{Umeda} \& {Nakamura}(2018)}]{umeda2018b}
{Umeda}, T., \& {Nakamura}, T. K.~M. 2018, Physics of Plasmas, 25, 102109,
  \dodoi{10.1063/1.5050542}

\bibitem[{{Umeda} {et~al.}(2009){Umeda}, {Yamao}, \& {Yamazaki}}]{Umeda2009}
{Umeda}, T., {Yamao}, M., \& {Yamazaki}, R. 2009, \apj, 695, 574,
  \dodoi{10.1088/0004-637X/695/1/574}

\bibitem[{{Wilson} {et~al.}(2007){Wilson}, {Cattell}, {Kellogg}, {Goetz},
  {Kersten}, {Hanson}, {MacGregor}, \& {Kasper}}]{Wilson2007}
{Wilson}, L.~B., I., {Cattell}, C., {Kellogg}, P.~J., {et~al.} 2007, \prl, 99,
  041101, \dodoi{10.1103/PhysRevLett.99.041101}

\bibitem[{{Wilson} {et~al.}(2010){Wilson}, {Cattell}, {Kellogg}, {Goetz},
  {Kersten}, {Kasper}, {Szabo}, \& {Wilber}}]{Wilson2010}
{Wilson}, L.~B., I., {Cattell}, C.~A., {Kellogg}, P.~J., {et~al.} 2010, \jgr,
  115, A12104, \dodoi{10.1029/2010JA015332}

\bibitem[{{Wu} \& {Lee}(1979)}]{Wu1979}
{Wu}, C.~S., \& {Lee}, L.~C. 1979, \apj, 230, 621, \dodoi{10.1086/157120}

\bibitem[{{Xie}(2019)}]{Xie2019}
{Xie}, H.~S. 2019, Computer Physics Communications, 244, 343,
  \dodoi{10.1016/j.cpc.2019.06.014}

\bibitem[{{Yang} {et~al.}(2016){Yang}, {Huang}, {Liu}, {Parks}, {Wang}, {Lu},
  \& {Hu}}]{Yang2016}
{Yang}, Z., {Huang}, C., {Liu}, Y.~D., {et~al.} 2016, \apjs, 225, 13,
  \dodoi{10.3847/0067-0049/225/1/13}

\bibitem[{{Yang} {et~al.}(2015){Yang}, {Liu}, {Richardson}, {Lu}, {Huang}, \&
  {Wang}}]{Yang2015}
{Yang}, Z., {Liu}, Y.~D., {Richardson}, J.~D., {et~al.} 2015, \apj, 809, 28,
  \dodoi{10.1088/0004-637X/809/1/28}

\bibitem[{{Yang} {et~al.}(2018){Yang}, {Lu}, {Liu}, \& {Wang}}]{Yang2018}
{Yang}, Z., {Lu}, Q., {Liu}, Y.~D., \& {Wang}, R. 2018, \apj, 857, 36,
  \dodoi{10.3847/1538-4357/aab714}

\bibitem[{{Yang} {et~al.}(2009){Yang}, {Lu}, {Lemb{\`e}ge}, \&
  {Wang}}]{Yang2009}
{Yang}, Z.~W., {Lu}, Q.~M., {Lemb{\`e}ge}, B., \& {Wang}, S. 2009, \jgr, 114,
  A03111, \dodoi{10.1029/2008JA013785}

\bibitem[{{Zank} {et~al.}(2006){Zank}, {Li}, {Florinski}, {Hu}, {Lario}, \&
  {Smith}}]{Zank2006}
{Zank}, G.~P., {Li}, G., {Florinski}, V., {et~al.} 2006, Journal of Geophysical
  Research (Space Physics), 111, A06108, \dodoi{10.1029/2005JA011524}

\bibitem[{{Zank} {et~al.}(1996){Zank}, {Pauls}, {Cairns}, \& {Webb}}]{Zank1996}
{Zank}, G.~P., {Pauls}, H.~L., {Cairns}, I.~H., \& {Webb}, G.~M. 1996, \jgr,
  101, 457, \dodoi{10.1029/95JA02860}

\bibitem[{{Zhao} {et~al.}(2019){Zhao}, {Liu}, {Hu}, \& {Wang}}]{Zhao2019}
{Zhao}, X., {Liu}, Y.~D., {Hu}, H., \& {Wang}, R. 2019, \apj, 882, 122,
  \dodoi{10.3847/1538-4357/ab379b}

\end{thebibliography}
\bibliographystyle{aasjournal}


\begin{deluxetable}{cccccccccccc}
\tablenum{1}
\tablecaption{Setups of species parameters for the linear analysis\label{tab:tab1}}
\tablehead{Run  & $B/B_0$ & $N/N_0$ & Species & $n\%$ & $\beta_{//}$ & $\beta_{\perp}$ & $V_{dx}/V_A$ & $V_{dy}/V_A$ & $V_{dz}/V_A$ & $V_{ring}/V_A$ & $M$}
\startdata
1 (Region A)    & 1.0 & 1.0 & Inc. H$^+$  & 83.5  & 0.1 & 0.25 & 9.00 & -0.16     & 0 & 0 & 0 \\
\               & \ & \     & Ref. H$^+$  & 16.5  & 0.1  & 0.75 & -1.60  & 14.8  & 0 & 0 & 0 \\
\               & \ & \     & Core e$^-$  & 90.0 & 0.3  & 0.75  & 7.25 & 2.30   & 0 & 0 & 1 \\
\               & \ & \     & Ring e$^-$  & 10.0 & 0.3  & 0.1  & 7.25 & 2.30   & 0 & 25.0 & 1 \\
\hline
2 (Region B)    & 2.0 & 2.0 & Inc. H$^+$      & 55.0  & 0.1  & 0.5  & 7.82  & -0.83 & 0 & 0    & 0 \\
\               & \ & \     & New ref. H$^+$  & 25.4  & 0.1  & 0.5  & -3.50 & 5.0   & 0 & 0    & 0 \\
\               & \ & \     & Old ref. H$^+$  & 19.6  & 0.1  & 0.5  & 3.63  & 12.5  & 0 & 0    & 0 \\
\               & \ & \     & Core e$^-$      & 90.0  & 0.3  & 2.0  & 4.12  & 3.26  & 0 & 0    & 1 \\
\               & \ & \     & Ring e$^-$      & 10.0  & 0.3  & 0.1  & 4.12  & 3.26  & 0 & 28.0 & 1 \\
\enddata
\tablecomments{(1) The species can be treated either magnetized ($M=1$) or unmagnetized ($M=0$). (2) $J=8$ is used for the $J-$pole Pad\'{e} expansion. (3) The electron VDF is contributed by a hot drifting Maxwellian core and a relatively cool Maxwellian ring. The ring velocity distribution and its drift across field are modeled as that used by \citet{umeda2012b} and \citet{umeda2018b}, respectively. The ion VDF is a superposition of two or three drifting Maxwellian subpopulations. As shown in Figure 2, ``Inc." and ``Ref." are abbreviations of ``incident" and ``reflected", respectively. ``New ref." represents ions which are newly reflected at the ramp and moving toward upstream. ``Old ref." refers to the ions that are reflected at earlier time, and they are gyrating back toward the downstream at this time. (4) $V_{dx,y,z}$ and $V_{ring}$ indicate the drift velocity components and the radius of modeled ring distributions. (5) We consider the nearly perpendicular wave modes. Wave normal angles (WNA) $\theta_{kB}=89.9^{\circ}$ and $89^{\circ}$ are employed for regions A and B, respectively. (6) All parameters are issued from the simulation and averaged over the sampling period in the shock ramp rest frame.}
\end{deluxetable}

\clearpage

\begin{figure}
\epsscale{1.05}
\plotone{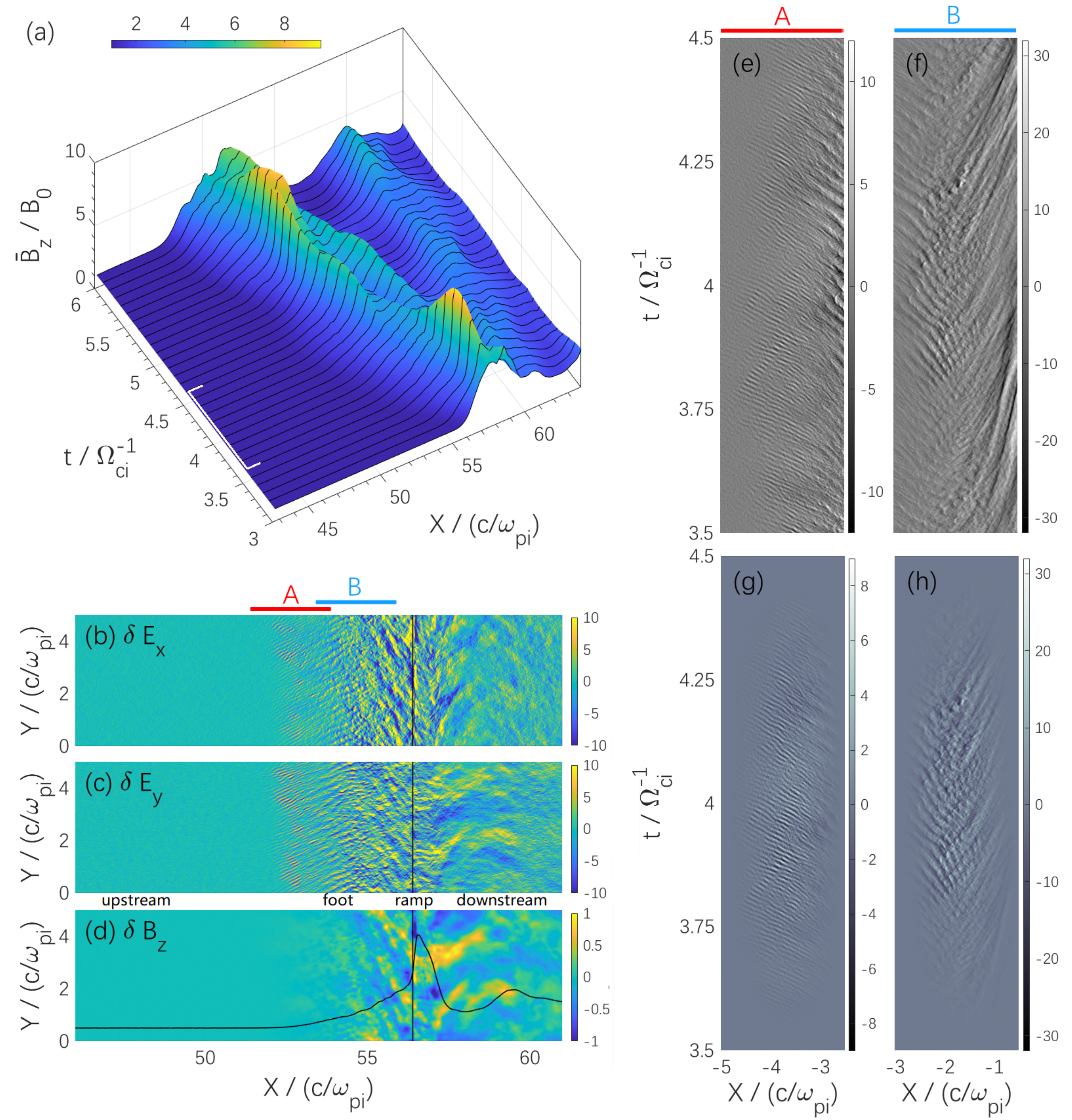}
\caption{\label{fig:fig1}{(a) An overview of the time-evolving $Y-$averaged shock magnetic field $\overline{B}_z$. The sampling period for the wave analysis is denoted by white lines. (b-d) Snapshots of electromagnetic fluctuations $\delta E_x$, $\delta E_y$ and $\delta B_z$ at $t=3.5\Omega_{ci}^{-1}$ within the sampling period. The ramp location $X=56.4d_i$ is marked by black vertical lines. The $Y-$averaged $B_z/2$ profile is shown in panel (d) for reference (black curve). Sampling regions A and B are marked out by red and blue horizontal lines on the top of panel (b). (e-f) Cross sections of 3-D sampled-data $\delta E_x(x,y,t)$ at $Y=L_y/2$ fron regions A and B, respectively. They are plotted in the shock rest frame where $X_{ramp}=0$. (g-h) Hanning windows are used to compensate for the nonperiodicity of the data, in both $X$ and $t$ directions. The same post-processing is performed on other fluctuation components before doing the 3-D FFT.}}
\end{figure}

\clearpage

\begin{figure}
\epsscale{1.15}
\plotone{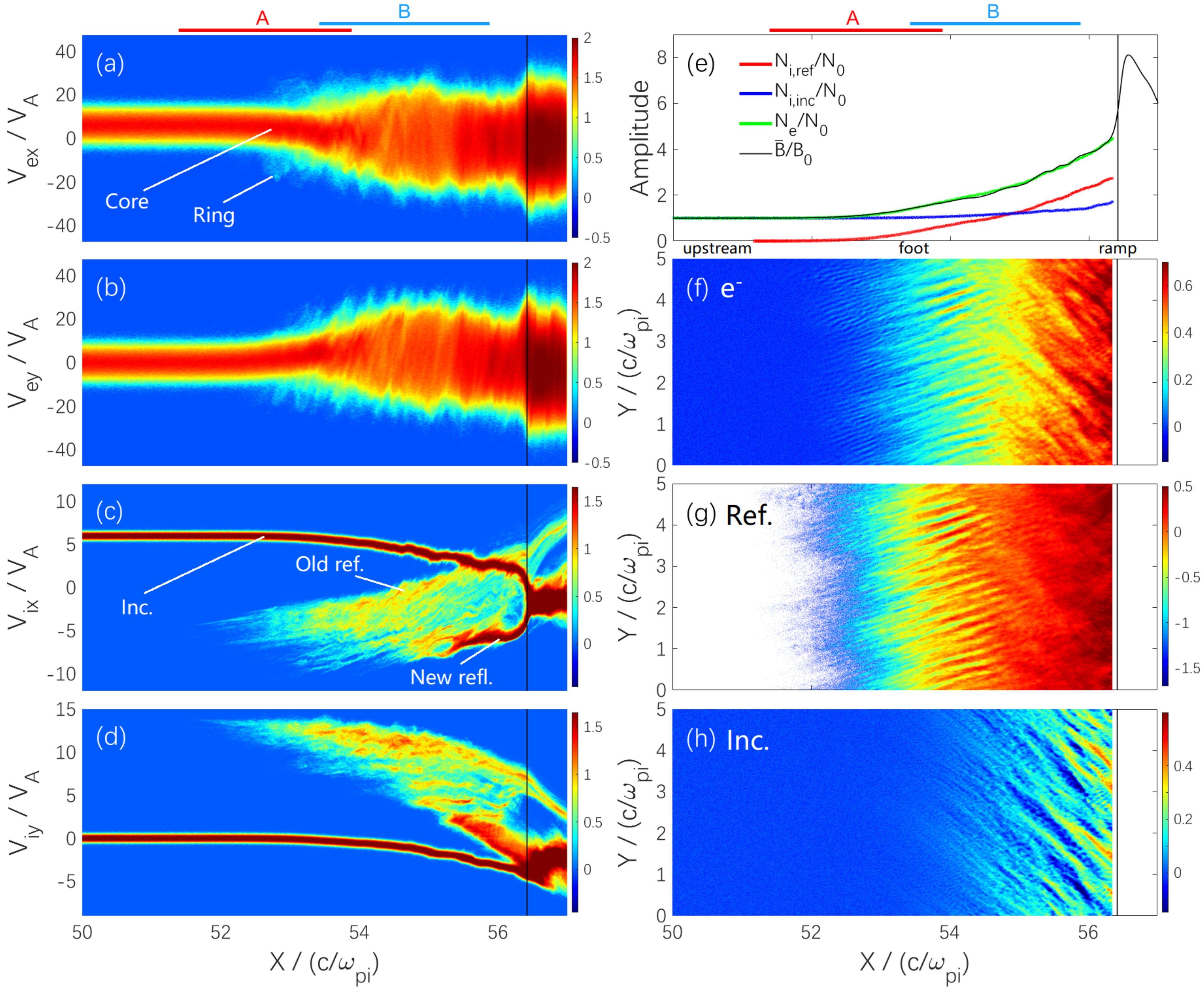}
\caption{\label{fig:fig2}{(a-b) Phase space plots $X-V_{ex,y}$ of electrons at $t=3.5\Omega_{ci}^{-1}$. The shock ramp location is marked by a vertical black line. Sampling regions A and B are denoted on the top. The core and ring components are marked in panel (a). In order to see these two components clearly, the particles located at $Y=L_y/2\pm0.1d_i$ are sampled for these plots. (c-d) Similar plots for ions. Incident ions (``Inc."), freshly reflected ions (``New ref.") and old gyro-reflected ions (``Old ref.") are marked in panel (c). (e) $Y-$averaged number density profiles of reflected ions (red), incident ions (blue), and electrons (green). The $Y-$averaged magnetic field $\overline{B}$ (black) is also shown for reference. Here, we only focus on the wave excitation region and separated the reflected ions ahead of the ramp as in \citet{Otsuka2019}. (f-h) Corresponding 2-D number density profiles.}}
\end{figure}

\clearpage

\begin{figure}
\epsscale{1.2}
\plotone{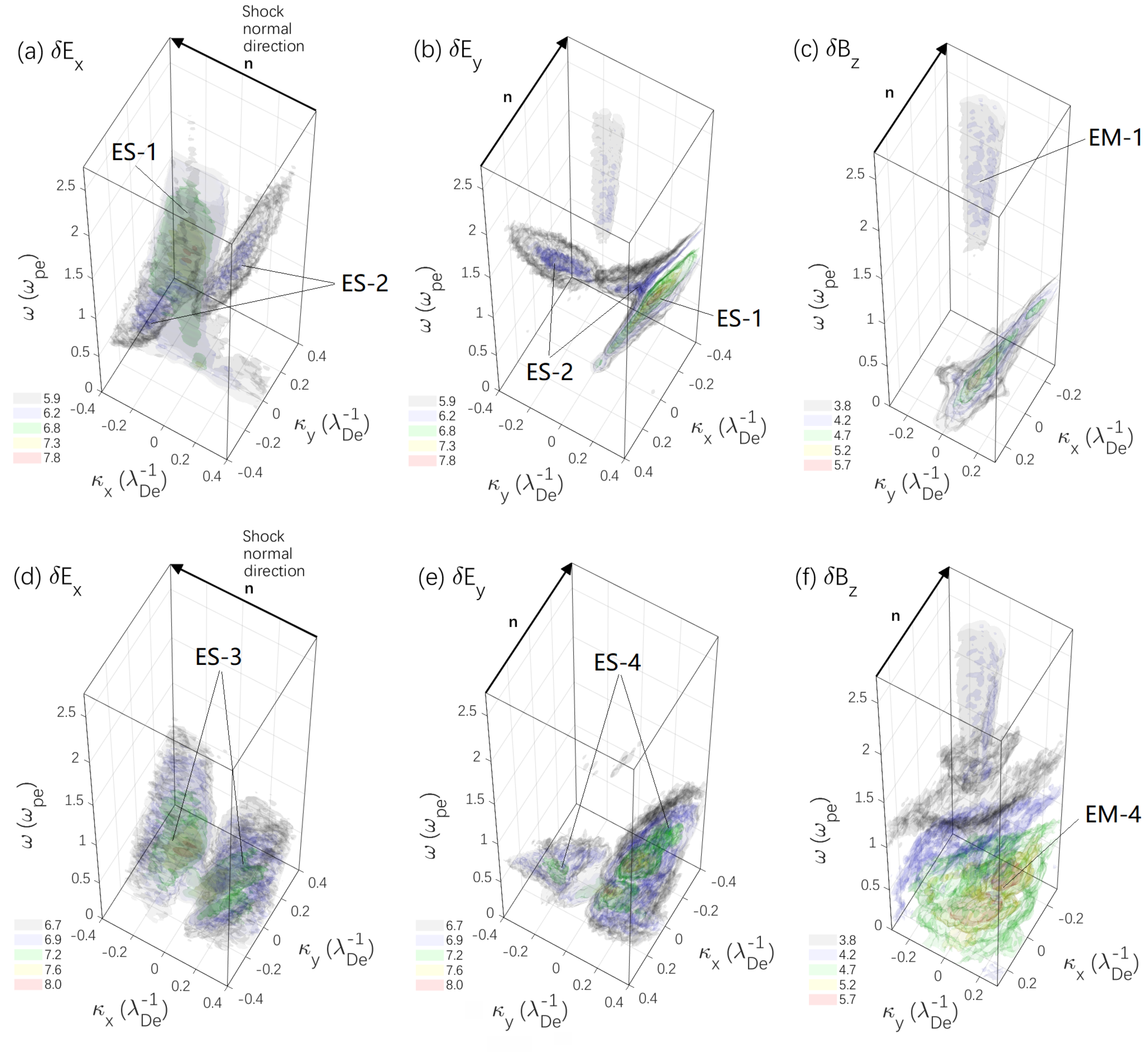}
\caption{\label{fig:fig3}{(a-c) A 3-D view of the dispersion relation diagram of fluctuating fields $\delta E_x$, $\delta E_y$ and $\delta B_z$ in region A of the shock front. The color indicates the amplitude of Fourier transformed fields in the $\kappa_x-\kappa_y-\omega$ space. The excited ES and EM waves are marked. (d-f) Similar plots as in (a-c) for fluctuating fields in region B. On the upper edge of each panel, the black arrow indicates the shock normal direction $\bf n$.}}
\end{figure}

\clearpage

\begin{figure}
\epsscale{1.0}
\plotone{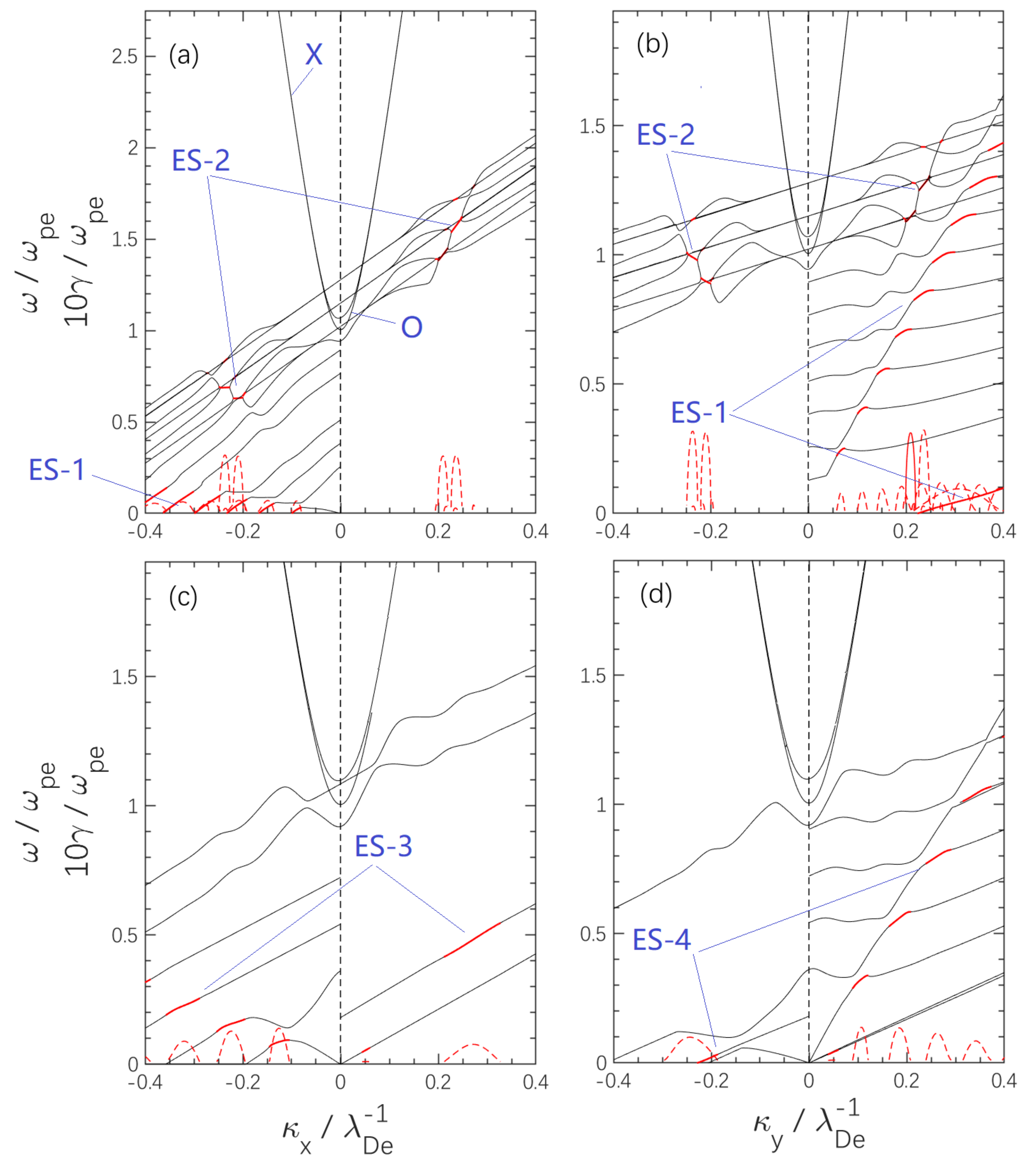}
\caption{\label{fig:fig4}{A unified numerically solvable framework for complicated kinetic plasma dispersion relations (pdrk) \citep{Xie2019} is used to identify the corresponding wave modes in Figure 3. Setups of the linear analysis are shown in Table 1. Real parts of main solutions are represented by black curves. Red line segments on these black curves indicate where the wave modes have positive growth rates. Corresponding zoomed imaginary parts are denoted by red dashed curves. Some high-frequency waves (e.g., O and X modes) are also plotted for reference in black. The labeled wave modes are described in the text.}}
\end{figure}



\end{document}